\documentstyle{article}
\begin{document}
\title {
\bf \huge A quantum-copying machine for equatorial qubits
}

\author{
{\bf
Heng Fan$^{a}$, Xiang-Bin Wang$^{b}$, Keiji Matsumoto$^{b,c}$}\\
\normalsize
$^{a}$Department of Physics, Graduate School of Science,\\
\normalsize University of Tokyo, Hongo 7-3-1,
Bunkyo-ku, Tokyo 113-0033, Japan.\\
\normalsize
$^{b}$ERATO, Japan Science and Technology Corporation,\\
\normalsize
Hongo 5-28-3, Bunkyo-ku, Tokyo 133-0033, Japan.\\
\normalsize
$^{c}$Department of Mathematical Engineering and Information Physics,\\
\normalsize
University of Tokyo, Hongo 7-3-1, Bunkyo-ku, Tokyo 113, Japan.
}
\maketitle

\begin{abstract}
Bu\v{z}ek and Hillery proposed a universal quantum-copying machine
(UQCM) (i.e., transformation)
to analyze the possibility of cloning arbitrary states. The UQCM copies
quantum-mechanical states with the quality of its output does not
depend on the input. We propose a slightly different transformation
to analyze a restricted set of input states.
We impose the conditions (I) the density matrices of the two
output states are the same, and that (II) the distance between
input density operator and the output density operators is input
state independent. 
Using Hilbert-Schmidt norm and Bures fidelity,
we show that our transformation can achieves the
bound of the fidelity.
\end{abstract}              
       

\baselineskip 0.5truecm
\newpage
\section{Introduction}
Quantum computing and quantum information have been attracting a great
deal of interests. They differ in many aspects from the classical
theories. One of the most fundamental differences between classical and
quantum information is the no-cloning theorem\cite{WZ}. It tells us that
arbitrary quantum information can not be copied exactly.
The no-cloning theorem for pure states is also extended to the case
that a general mixed state can not be broadcast\cite{BCFJ}. 
However, no-cloning theorem does not forbid imperfect cloning.
And several kinds of quantum copying (cloning) machines (QCM)
are proposed\cite{BH,BDEF,F}.
Some authors also try to find the optimal QCMs\cite{BDEF,BEM,GM,W}.

In the proof of the no-cloning theorem, Wootters and Zurek introduced
a QCM which has the property that the
quality of the copy it makes depends on the input states\cite{WZ}.
To diminish or cancel this disadvantage, Bu\v{z}ek and Hillery proposed a
UQCM, the copying process is input-state independent.
They use Hilbert-Schmidt norm to quantify distance between
input density operator and the output density operators.
Bru\ss ~et al discussed the performance of a UQCM by
analyzing the role of the symmetry and isotropy conditions
imposed on the system and found the optimal UQCM and the
optimal state-dependent quantum cloning\cite{BDEF}.
Optimal quantum cloning of general $N\rightarrow M$ case is
discussed in Ref.\cite{BEM,GM,W}.
The relation between quantum cloning and superluminal
signalling is proposed and discussed in Ref.\cite{G,BDMS}.

In this paper, we propose a QCM for a restricted set of input
states. The Bloch vector is restricted to the intersection
of $x-z$ ($x-y$ and $y-z$) plane with the Bloch sphere, this
kind of qubits are the so-called equatorial qubits\cite{BCDM}.
Applying the
method by Bu\v{z}ek and Hillery, we propose a possible extension of
the original transformation. We demand that (I) the density matrices of
the two output states are the same, and that (II) the distance between
input density operator and the output density operators is input
state independent. To evaluate the distance of two states, we use
both Hilbert-Schmidt norm and Bures fidelity.
There are a family of transformations which satisfy the above
two conditions. In a special point, we can obtain an optimal
fidelity and the correspondent transformation agrees with
the results of Bru\ss ~et al \cite{BCDM} who study the
optimal quantum cloning for equatorial qubits.
The fidelity of quantum cloning for the equatorial qubits
is higher than the original Bu\v{z}ek and Hillery UQCM.
This is expected as that the more information about the input
is given, the better one can clone each of its states.
We also obtain the quantum cloning transformations
for equatorial qubits in $x-y$ and $y-z$ planes.

The paper is organized as follows: In Section 2,
we introduce the transformation for equator in
$x-z$ plane. In Section 3, we use
Hilbert-Schmidt norm to evaluate the distance between input
state and output states, and the minimal distance
is found. In section 4, we use Bures metric to define the
fidelity, and the condition of orientation invariance
of the Bloch vector is studied. In section 5, the
cloning transformations for equator in $x-y$ and $y-z$
planes are obtained.
Section 6 includes summary and discussions.

\section{Transformation}
We propose the following transformation
\begin{eqnarray}
|0>_a|Q>_x\rightarrow
\left( |0>_a|0>_b+\lambda |1>_a|1>_b\right) |Q_0>_x
+\left( |1>_a|0>_b+|0>_a|1>_b\right) |Y_0>_x, \nonumber \\
|1>_a|Q>_x\rightarrow
\left( |1>_a|1>_b+\lambda |0>_a|0>_b\right) |Q_1>_x
+\left( |1>_a|0>_b+|0>_a|1>_b\right) |Y_1>_x,
\label{tran}
\end{eqnarray}
where the states $|Q_j>_x, |Y_j>_x, j=0,1$ are not necessarily
orthonormal. Hereafter, we will drop the subscript $x$ for convenience.
Explicitly, this transformation is a generalization of the original
one proposed by Bu\v{z}ek and Hillery\cite{BH}.
When $\lambda =0$, this transformation is reduced to the original
transformation. For convenience, we restrict $\lambda $ to be real
and $\lambda \not=\pm 1$. We also assume
\begin{eqnarray}
<Q_0|Q_1>=<Q_1|Q_0>=0.
\end{eqnarray}
Considering the unitarity of the transformation,
we have the following relations:
\begin{eqnarray}
&&(1+\lambda ^2)<Q_j|Q_j>+2<Y_j|Y_j>=1,~~ j=0,1,
\\
&&<Y_0|Y_1>=<Y_1|Y_0>=0.
\end{eqnarray}
As proposed by Bu\v{z}ek and Hillery, we further assume the following
relations to reduce the free parameters
\begin{eqnarray}
&&<Q_j|Y_j>=0, ~~j=0,1.\\
&&<Y_0|Y_0>=<Y_1|Y_1>\equiv \xi,\\
&&<Y_0|Q_1>=<Q_0|Y_1>
=<Q_1|Y_0>=<Y_1|Q_0>\equiv {\eta \over 2}.
\end{eqnarray}
For simplicity, we will also use the following standard notations
\begin{eqnarray}
|jk>=|j>_a|k>_b, ~~j,k=0,1,
\end{eqnarray}
and
\begin{eqnarray}
|+>=\frac 1{\sqrt{2}}(|10>+|01>), ~~
|->=\frac 1{\sqrt{2}}(|10>-|01>).
\label{base}
\end{eqnarray}
Obviously, $|\pm >$ and $|00>, |11>$ constitute an orthonormal basis. 
The input state is a pure superposition state (SU(2) coherent state)
\begin{eqnarray}
|\psi >_a=\alpha |0>_a+\beta |1>_a
\label{input}
\end{eqnarray}
with $\alpha ^2+\beta ^2=1$. Here, we use an assumption that
$\alpha $ and $\beta $ are real that means the $y$ component
of the Bloch vector of the qubits is zero. And we
just deal with a restricted set of input states.

The output density operator $\rho _{ab}^{(out)}$ describing
output state after the copying procedure reads
\begin{eqnarray}
\rho _{ab}^{(out)}&=&|00><00|\{
\frac {1-2\xi }{1+\lambda ^2}[\lambda ^2+\alpha ^2(1-\lambda ^2)]\}
\nonumber \\
&&+(|00><10|+|00><01|+|11><10|+|11><01|
+|01><00|+|10><00|
\nonumber \\
&&+|01><11|+|10><11|)
[{\eta \over 2}\alpha \beta (\lambda +1)]
\nonumber \\
&&+(|00><11|+|11><00|)(\frac {1-2\xi }{1+\lambda ^2}\lambda )
\nonumber \\
&&+\xi (|01><10|+|01><01|+|10><10|+|10><01|)
\nonumber \\
&&+|11><11|\{ \frac {1-2\xi }{1+\lambda ^2}[\alpha ^2(\lambda ^2-1)+1]\},
\end{eqnarray}
where 
$\rho _{ab}^{(out)}=Tr_x[\rho _{abx}^{(out)}]$ with
$\rho _{abx}^{(out)}\equiv |\Psi >_{abx}^{(out)}
{_{abx}^{(out)}<\Psi |}$. Taking trace on mode $b$ or mode $a$,
we can get reduced density operator for mode $a$ or mode $b$,
$\rho _a^{(out)}$ or $\rho _b^{(out)}$,
\begin{eqnarray}
\rho _a^{(out)}=\rho _b^{(out)}&=&|0><0|
\left( (\alpha ^2+\lambda ^2\beta ^2)\frac {1-2\xi }{1+\lambda ^2}+\xi\right)
\nonumber \\
&&(|0><1|+|1><0|)\alpha \beta \eta (1+\lambda )
+|1><1|\left(
\xi +(\beta ^2+\lambda ^2\alpha ^2)\frac {1-2\xi }{1+\lambda ^2} \right).
\label{Pa}
\end{eqnarray}
The density operators $\rho _a^{(out)}$ and $\rho _b^{(out)}$ are exactly
the same. We see that the output density operators are identical to
each other. However, it is well known that
they are not equal to the original input density operator.
Next, we first use Hilbert-Schmidt norm to evaluate the distance between
input density operator and output density operators.

\section{Hilbert-Schmidt norm}
For two-dimensional space, the Hilbert-Schmidt norm is believed to 
give a reasonable result in comparing density matrices though
it becomes less good for finite-dimensional spaces
as the dimension increases.
The Hilbert-Schmidt norm define the distance between input density
operator and output density operator as
\begin{eqnarray}
D_a\equiv Tr[\rho _a^{(out)}-\rho _a^{(in)}]^2,
\end{eqnarray}
where $\rho _a^{(in)}$ is the input density operator.
The distance between the two-mode density operators
$\rho _{ab}^{(out)}$ and
$\rho _{ab}^{(in)}=\rho _a^{(in)}\otimes \rho _a^{(in)}$ is defined
as:
\begin{eqnarray}
D_{ab}^{(2)}=Tr[\rho _{ab}^{(out)}-\rho _{ab}^{(in)}]^2.
\label{d2}
\end{eqnarray}
With the help of relation (\ref{Pa}), we find
\begin{eqnarray}
D_a&=&\{ \xi +\frac {1-2\xi }{1+\lambda ^2}[\alpha ^2(1-\lambda ^2)
+\lambda ^2]-\alpha ^2\} ^2
+2\alpha ^2(1-\alpha ^2)(\lambda \eta +\eta -1)^2
\nonumber \\
&&+\{ \xi -1+\frac {1-2\xi }{1+\lambda ^2}[1+\alpha ^2(\lambda ^2-1)]
+\alpha ^2\} ^2.
\end{eqnarray}
We demand that this distance is independent of the parameter
$\alpha ^2$. That means the quality of the copies it makes
is independent of the input state.
\begin{eqnarray}
\frac {\partial} {\partial \alpha ^2}D_a=0
\end{eqnarray}
We can choose the following solution
\begin{eqnarray}
\eta =\frac {1-\lambda }{1+\lambda ^2}(1-2\xi ).
\label {eta}
\end{eqnarray}
Thus we get
\begin{eqnarray}
D_a=2\left( \xi \frac {1-\lambda ^2}{1+\lambda ^2}
+\frac {\lambda ^2}{1+\lambda ^2}\right) ^2.
\end{eqnarray}
In case $\lambda =0 $, we find
$\eta =1-2\xi $ and $D_a=2\xi ^2$. These are exactly the original results
obtained by Bu\v{z}ek and Hillery \cite{BH}.

To find the result of $D_{ab}^{(2)}$, we can rewrite the output density
operator $\rho _{ab}^{(out)}$ by choose basis in (\ref{base}).
Substituting the relation (\ref{eta}) into the two-mode output
density operator, we can obtain
\begin{eqnarray}
\rho _{ab}^{(out)}&=&
|00><00|\{ \frac {1-2\xi }{1+\lambda ^2}[\lambda ^2
+\alpha ^2(1-\lambda ^2)]\}
\nonumber \\
&&+\left( |00><+|+|+><00|
+|11><+|+|+><11|\right)
\{ \sqrt{2}\alpha \beta \frac {1-\lambda ^2}{2(1+\lambda ^2)}
(1-2\xi )\}
\nonumber \\
&&+\left( |00><11|+|11><00| \right)
\{ \frac {1-2\xi }{1+\lambda ^2}\lambda \}
\nonumber \\
&&+2\xi |+><+|+|11><11|\{ \frac {1-2\xi }{1+\lambda ^2}
[\alpha ^2(\lambda ^2-1)+1]\}.
\end{eqnarray}
By straightforward calculations, we can write
\begin{eqnarray}
\rho _{ab}^{(in)}&=&\alpha ^4|00><00|+
\sqrt{2}\alpha ^3\beta(|00><+|+|+><00|)
+\alpha ^2\beta ^2(|00><11|+|11><00|)
\nonumber \\
&&+2\alpha ^2\beta ^2|+><+|
+\sqrt{2}\alpha \beta ^3(|+><11|+|11><+|)
+\beta ^4|11><11|.
\end{eqnarray}
And with the definition (\ref{d2}), we have
\begin{eqnarray}
D_{ab}^{(2)}=(U_{11})^2 +(U_{22})^2+(U_{33})^2
+2(U_{12})^2+2(U_{13})^2+2(U_{23})^2,
\end{eqnarray}
where
\begin{eqnarray}
&&U_{11}=\alpha ^4-\frac {1-2\xi }{1+\lambda ^2}[\lambda ^2
+\alpha ^2(1-\lambda ^2)],
~~~U_{22}=2\xi -2\alpha ^2+2\alpha ^4,\nonumber \\
&&U_{33}=\alpha ^4-2\alpha ^2+1-
\frac {1-2\xi }{1+\lambda ^2}[\alpha ^2(\lambda ^2-1)+1],
~~~U_{12}=\sqrt{2}\alpha \beta [\alpha ^2-\frac {1-\lambda ^2}{1+\lambda ^2}
({1\over 2}-\xi )],
\nonumber \\
&&U_{13}=\alpha ^2\beta ^2-\frac {1-2\xi }{1+\lambda ^2}\lambda,
~~~U_{23}=\sqrt{2}\alpha \beta [\beta ^2-
\frac {1-\lambda ^2}{1+\lambda ^2}({1\over 2}-\xi )].
\end{eqnarray}
We still impose the condition
\begin{eqnarray}
\frac {\partial}{\partial \alpha ^2}D_{ab}^{(2)}=0.
\end{eqnarray}
We find the result
\begin{eqnarray}
\xi =\frac {(1-\lambda )^2}{2(3-2\lambda +3\lambda ^2)}.
\label{xi}
\end{eqnarray}
Substitute these results into $D_a$ and $D_{ab}^{(2)}$, we have
\begin{eqnarray}
D_a&=&\frac {(1-2\lambda +5\lambda ^2)^2}
{2(3-2\lambda +3\lambda ^2)^2},
\nonumber \\
D_{ab}^{(2)}&=&\frac {2(1-4\lambda +12\lambda ^2-8\lambda ^3+7\lambda ^4)}
{(3-2\lambda +3\lambda ^2)^2}.
\label{distance}
\end{eqnarray}
So, we actually can have a family of transformations
to satisfy the two conditions (I) and (II). 
In case $\lambda =0$, we have Bu\v{z}ek and Hillery's result
\begin{eqnarray}
D_a=\frac {1}{18}\approx 0.056,
~~~D_{ab}^{(2)}=\frac {2}{9}\approx 0.22.
\end{eqnarray}
Our aim is to find smaller $D_a$ and $D_{ab}^{(2)}$ for
equatorial qubits. We can calculate that in the
region $0<\lambda <1/3$, both
$D_a$ and $D_{ab}^{(2)}$ take smaller values than
the case $\lambda =0$. When we choose
\begin{eqnarray}
\lambda =3-2\sqrt{2},
\label{lambda}
\end{eqnarray}
both $D_a$ and $D_{ab}^{(2)}$ take their minimal values,
\begin{eqnarray}
D_a=\frac {99-70\sqrt{2}}{68-48\sqrt{2}}
\approx 0.043,
~~~D_{ab}=\frac {215-152\sqrt{2}}{8(3-2\sqrt{2})^2}
\approx 0.17.
\end{eqnarray}
Thus for equatorial qubits,
we can find smaller $D_a$ and $D_{ab}^{(2)}$, that means
this QCM (\ref{tran}) has a higher fidelity than
the original UQCM \cite{BH} by using the Hilbert-Schmidt norm.
Actually, because that we assume $\alpha $ and $\beta $ are real,
only a single unknown parameter is copied instead of two
unknown parameters for the case of a general pure state.
Thus a higher fidelity of quantum cloning can be achieved.
The case of spin flip has a similar phenomenon\cite{BHW, BBHB, B}. 

Under the condition (\ref{lambda}), we have
\begin{eqnarray}
\xi =\frac {1}{8},
~~~\eta =\frac {\sqrt{2}-1}{12-8\sqrt{2}}.
\end{eqnarray}
We can realize vectors $|Q_j>, |Y_j>, j=0,1$ in two-dimensional space
\begin{eqnarray}
&&|Q_0>=(0, \frac {1}{4-2\sqrt{2}}), ~~~
|Q_1>=(\frac {1}{4-2\sqrt{2}},0),\nonumber \\
&&|Y_0>=(\frac {1}{2\sqrt{2}}, 0),~~~
|Y_1>=(0, \frac {1}{2\sqrt{2}}).
\end{eqnarray}
The transformation (\ref{tran}) can be rewritten as
\begin{eqnarray}
|0>_a|Q>_x&\rightarrow &
\frac {1}{4-2\sqrt{2}}[|00>+(3-2\sqrt{2})|11>]|\uparrow >
+\frac 12|+>|\downarrow >,\\
|1>_a|Q>_x&\rightarrow &
\frac {1}{4-2\sqrt{2}}[|11>+(3-2\sqrt{2})|00>]|\downarrow >
+\frac 12|+>|\uparrow >.
\end{eqnarray}
This transformations agree with the results obtained by
Bru\ss~et al\cite{BCDM}.

For an arbitrary $\lambda $ with the condition
(\ref{eta}) and (\ref{xi}) satisfied, we can still
realize vectors $|Q_j>$, $|Y_j>$, $j=0,1$ in two-dimensional space,
\begin{eqnarray}
|Q_0>&=&q|\uparrow >,
~~~|Q_1>=q|\downarrow >,
\nonumber \\
|Y_0>&=&y|\downarrow >,
~~~|Y_1>=y|\uparrow >,
\label{twod}
\end{eqnarray}
where we use notations
\begin{eqnarray}
q\equiv \sqrt{\frac {2}{3-2\lambda +3\lambda ^2}},
~~~y\equiv \frac {1-\lambda }{\sqrt{6-4\lambda +6\lambda ^2}}.
\label{qy}
\end{eqnarray}
Thus all transformations (\ref{tran}) satisfy the
condition (I) and (II). Explicitly, the quantum cloning
transformation for pure input states (\ref{input}) can be written as
\begin{eqnarray}
|0>|Q>_x\rightarrow
\left( |00>+\lambda |11>\right)q|\uparrow >_x
+\left( |10>+|01>\right) y|\downarrow >_x, \nonumber \\
|1>|Q>_x\rightarrow
\left( |11>+\lambda |00>\right) q|\downarrow >_x
+\left( |10>+|01>\right) y|\downarrow >_x.
\label{tran2}
\end{eqnarray}
The distances defined by Hilbert-Schmidt norm take the
form (\ref{distance}).

\section{Bures fidelity}
For finite-dimensional spaces, Hilbert-Schmidt norm becomes less
good when the dimension increases. Bures fidelity provides a
more exact measurement of the distinguishability of two density
matrices. In this section, we will use Bures fidelity to check
the result in the previous section.
The fidelity is defined as
\begin{eqnarray}
F(\rho _1, \rho _2)=Tr(\rho _1^{1/2}\rho _2\rho _1^{1/2})^{1/2}.
\end{eqnarray}
The values of $F$ range from 0 to 1, a larger F corresponds to a higher
fidelity. $F=1$ means two density matrices are equal.

We have a matrix
\begin{eqnarray}
U=\left( \begin{array}{cc}
-\frac {\beta }{\alpha }& \frac {\alpha }{\beta }\\
1&1\end{array}\right)
\end{eqnarray}
to diagonalize $\rho _a^{(in)}$\cite{KOWY}
\begin{eqnarray}  
\rho _a^{(in)}=U\left( \begin{array}{cc} 0&0\\
0&1\end{array}\right) U^{-1}.
\end{eqnarray}
We thus have
\begin{eqnarray}
F(\rho _a^{(in)}, \rho _a^{(out)})=
\{ \xi
+\frac {(1-2\xi )[2\alpha ^4(1-\lambda ^2)
+2\alpha ^2(\lambda ^2-1)+1]}{1+\lambda ^2}
+2\alpha ^2(1-\alpha ^2)\eta (\lambda +1)\} ^{1/2},
\end{eqnarray}
We demand that the fidelity be independent of the input state
\begin{eqnarray}
\frac {\partial }{\partial \alpha ^2}
F(\rho _a^{(in)}, \rho _a^{(out)})=0,
\end{eqnarray}
we can find
\begin{eqnarray}
\eta =\frac {1-\lambda }{1+\lambda ^2}(1-2\xi ),
\label{eta1}
\end{eqnarray}
with
\begin{eqnarray}
F(\rho _a^{(in)}, \rho _a^{(out)})=\left(
\frac {1-\xi +\lambda ^2\xi }{1+\lambda ^2}\right) ^{1/2}.
\end{eqnarray}
Next, we use Bures fidelity to evaluate the distinguishability of
density operators $\rho _{ab}^{(out)}$ and
$\rho _{ab}^{(in)}=\rho _a^{(in)}\otimes \rho _a^{(in)}$.
We have
\begin{eqnarray}
F(\rho _{ab}^{(in)}, \rho _{ab}^{(out)})
&=&\{ \frac {1-2\xi }{1+\lambda ^2}[\lambda ^2+
\alpha ^2(1-\lambda ^2)]\alpha ^4
+2\alpha ^2(1-\alpha ^2)\lambda \frac {1-2\xi }{1+\lambda ^2}
+2\alpha ^2(1-\alpha ^2)(1-2\xi )\frac {1-\lambda ^2}{1+\lambda ^2}
\nonumber \\
&&+4\alpha ^2(1-\alpha ^2)\xi
+\frac {1-2\xi }{1+\lambda ^2}
[\alpha ^2(\lambda ^2-1)+1](1-\alpha ^2)^2\} ^{1/2}.
\end{eqnarray}
We still impose the condition
\begin{eqnarray}
\frac {\partial }{\partial \alpha ^2}
F(\rho _{ab}^{(in)}, \rho _{ab}^{(out)})=0.
\end{eqnarray}
We have
\begin{eqnarray}
\xi =\frac {(1-\lambda )^2}{2(3-2\lambda +3\lambda ^2)}.
\label{xi1}
\end{eqnarray}
Thus, we finally have two Bures fidelities for one and two-mode
density operators,
\begin{eqnarray}
F(\rho _a^{(in)}, \rho _a^{(out)})&=&
\left[ \frac {5-2\lambda +\lambda ^2}{2(3-2\lambda +3\lambda ^2)}
\right] ^{1/2},
\label{f1} \\
F(\rho _{ab}^{(in)}, \rho _{ab}^{(out)})
&=&\left[ \frac {2}{3-2\lambda +3\lambda ^2}\right] ^{1/2}.
\label{f2}
\end{eqnarray}
We can find that for both Hilbert-Schmidt norm and
Bures fidelity, we have the same relations (\ref{eta}, \ref{eta1}) and
(\ref{xi},\ref{xi1}). However, the fidelity (\ref{f1}) and
(\ref{f2}) do not take the maximums simultaneously which is
different from the case of Hilbert-Schmidt norm.
$F(\rho _a^{(in)}, \rho _a^{(out)})$ takes its maximum when
$\lambda =3-2\sqrt{2}$ the same as the case of Hilbert-Schmidt
norm.
$F(\rho _{ab}^{(in)}, \rho _{ab}^{(out)})$ takes its maximum when
$\lambda =1/3$ which is different from the case of
Hilbert-Schmidt norm. In the region $0<\lambda <1/3$,
for both
$F(\rho _{ab}^{(in)}, \rho _{ab}^{(out)})$ and
$F(\rho _a^{(in)}, \rho _a^{(out)})$, we can have a higher
fidelity than the original UQCM which corresponds to $\lambda =0$,
this result agree with the previous result by Hilbert-Schmidt norm
in last section.
Here we remark that we just deal with the equatorial qubits.

When $\lambda =3-2\sqrt{2}$, $F(\rho _a^{(in)}, \rho _a^{(out)})$
takes its maximum
\begin{eqnarray}
F(\rho _a^{(in)}, \rho _a^{(out)})|_{\lambda =3-2\sqrt{2}}
=\left( \frac {2-\sqrt{2}}{12-8\sqrt{2}}\right) ^{1/2}
\approx 0.92388
\label{hfidelity}
\end{eqnarray}
which is larger than the original UQCM
\begin{eqnarray}
F(\rho _a^{(in)}, \rho _a^{(out)})|_{\lambda =0}
=\left( \frac {5}{6}\right) ^{1/2}
\approx 0.912871.
\end{eqnarray}
And we also have
\begin{eqnarray}
F(\rho _{ab}^{(in)}, \rho _{ab}^{(out)})|_{\lambda =3-2\sqrt{2}}
=\sqrt{\frac {1}{24-16\sqrt{2}}}
\approx 0.853553 >
F(\rho _{ab}^{(in)}, \rho _{ab}^{(out)})|_{\lambda =0}
=\sqrt{\frac {2}{3}}\approx 0.816497
\end{eqnarray}
Here the optimal fidelity (\ref{hfidelity}) also
agrees with the result obtained by Bru\ss ~et al\cite{BCDM}.

In studying the optimal UQCM, the condition of orientation invariance
of Bloch vector is generally imposed\cite{BDEF}. Under the symmetry
condition (I), the condition of orientation invariance of Bloch vector
is equivalent to the condition (II) that the distance between input
density operator and the output density operators is input state
independent. We can check that for the case under consideration
in this paper, the orientation invariance of Bloch vector
means the relation (\ref{eta}) or (\ref{eta1}) which is
the subsequence of condition (II).

\section{Quantum copying-machine for $x-y$ and $y-z$ planes
equatorial qubits}
In this section, instead of $x-z$ plane equator, we first study
the transformations for equatorial qubits in $x-y$ plane.
The input pure states take the form
\begin{eqnarray}
|\psi >=\frac {1}{\sqrt{2}}(|0>+e^{i\phi }|1>),
\label{input1}
\end{eqnarray}
where $\phi \in [0,2\pi )$. We can check that the $z$ component
of the Bloch vector is zero. Actually, the optimality of the
fidelity should be independent from the choice of a particular
basis. So, the optimal fidelity (\ref{hfidelity}) for
$x-z$ plane equator remains the
same for $x-y$ equator.
However, for different input states, the optimal
transformation generally should be different.
For $x-z$ plane equator,
we have already found that a family of transformations (\ref{tran})
with restrictions (\ref{twod}) satisfy the conditions of
quantum copying-machine (I) and (II), and also with the property
of orientation invariance of Bloch vector. We expect the same result
for the $x-y$ plane equator.

We can rewrite the input states as
\begin{eqnarray}
|\psi >&=&e^{\frac {i\phi }{2}}\frac {1}
{\sqrt{2}}\left( e^{-{\frac {i\phi }{2}}}|0>+
e^{{\frac {i\phi }{2}}}|1>\right) \nonumber \\ 
&=&
e^{\frac {i\phi }{2}}\frac {1}
{\sqrt{2}}[\cos {\frac {\phi }{2}} (|0>+|1>)
+i\sin {\frac {\phi }{2}}(|1>-|0>)],
\end{eqnarray}
It is obvious that two vectors $(|1>+|0>)/\sqrt{2}$
and $(|1>-|0>)/\sqrt{2}$ constitute an orthonormal basis.
And the input states of $x-y$ (\ref{input1}) equator can be redefined
as the input states of $x-z$ plane equator (\ref{input}).
With the help of results for $x-z$ plane equator,
we can calculate the quantum cloning transformation for
input states (\ref{input1}) as
\begin{eqnarray}
|0>|Q>_x\rightarrow
|00>\frac {2(1-\lambda )}{\sqrt{6-4\lambda +6\lambda ^2}}
|\uparrow >_x +\left( |01>+|10>\right)
\frac {1+\lambda }{\sqrt{6-4\lambda +6\lambda ^2}}|\downarrow >_x,
\\
|1>|Q>_x\rightarrow
|11>\frac {2(1-\lambda )}{\sqrt{6-4\lambda +6\lambda ^2}}
|\downarrow >_x +\left( |01>+|10>\right)
\frac {1+\lambda }{\sqrt{6-4\lambda +6\lambda ^2}}|\uparrow >_x.
\end{eqnarray}
The Bures fidelity takes the same value as the case of
$x-z$ equator (\ref{f1}).
When $\lambda =0$, we still obtain the results
of UQCM\cite{BH}.
When $\lambda =3-2\sqrt{2}$, we obtain the optimal
fidelity (\ref{hfidelity}). And the optimal quantum cloning transformation
for (\ref{input1}) becomes as
\begin{eqnarray}
|0>|Q>_x\rightarrow
\frac {1}{\sqrt{2}}|00>|\uparrow >_x
+{\frac 12}\left( |01>+|10>\right) |\downarrow >_x,
\\
|1>|Q>_x\rightarrow
\frac {1}{\sqrt{2}}|11>|\downarrow >_x
+{\frac 12}\left( |01>+|10>\right) |\uparrow >_x.
\end{eqnarray}

For the case of $y-z$ equator, the results are
similar as the case of $x-z$ plane. We can actually obtain
the results by rename some vectors. We consider the
input equatorial states as
\begin{eqnarray}
|\psi >=\cos {\theta }|0>+i\sin {\theta }|1>.
\end{eqnarray}
The general and the optimal quantum cloning transformations
can be written as follows:
\begin{eqnarray}
|0>|Q>_x\rightarrow
\left( |00>-\lambda |11>\right)q|\uparrow >_x
+\left( |10>+|01>\right) y|\downarrow >_x, \nonumber \\
|1>|Q>_x\rightarrow
\left( |11>-\lambda |00>\right) q|\downarrow >_x
+\left( |10>+|01>\right) y|\downarrow >_x,
\end{eqnarray}
where $q,y$ is defined in (\ref{qy}), and
\begin{eqnarray}
|0>_a|Q>_x&\rightarrow &
\frac {1}{4-2\sqrt{2}}[|00>-(3-2\sqrt{2})|11>]|\uparrow >
+\frac 12|+>|\downarrow >,\\
|1>_a|Q>_x&\rightarrow &
\frac {1}{4-2\sqrt{2}}[|11>-(3-2\sqrt{2})|00>]|\downarrow >
+\frac 12|+>|\uparrow >.
\end{eqnarray}

\section{Summary and discussions}
We propose QCMs for equatorial qubits with the
equator in the $x-z$, $x-y$ and $y-z$ planes respectively. We use both
Hilbert-Schmidt norm and Bures fidelity to define the
distinguishability of the density operator matrices.
We can have a family of transformations,
using Hilbert-Schmidt norm, the distances achieve
the minimal values simultaneously for both one
and two-mode operators in a special point $\lambda =3-2\sqrt{2}$.
Using Bures fidelity, the fidelity for one-mode
operators also achieves the bound of fidelity in the
case $\lambda =3-2\sqrt{2}$.

We use only two conditions, (I)two output density matrices
are identical and (II)the distance defined by Hilbert-Schmidt
norm and the Bures fidelity for input density operator matrix and
output density operator matrices are independent of the input
state. We checked that the system also has the
property of orientation
invariance of the Bloch vector which is generally imposed
to UQCM.
For the case of arbitrary input
state, $\lambda \not= 0$ generally breaks the condition
of orientation invariance of the Bloch vector.

When we use Bures fidelity, the fidelity for one and two-mode
does not reach their maximal points simultaneously which
is different from the case of Hilbert-Schmidt norm, it is still
need to clarify which measurement is more reasonable for
two-dimensional space. We only consider the
$1\rightarrow 2$ cloning transformation in this paper,
it is interesting to study the general $N\rightarrow M$ case.
And the case of mixed states is also worth studying.

\vskip 1truecm
\noindent
{\bf Acknowlegements}:
One of the authors HF acknowleges the support of JSPS and
the hospitality of Wadati group in Department of Physics,
University of Tokyo.  We thank V.Bu\v{z}ek for very
useful comments, and we thank D.Bru\ss ,
N.Gisin and M.Hillery for
communications.


\begin{thebibliography}{99}
\bibitem{WZ}W.K.Wootters, and W.H.Zurek, Nature (London){\bf 299},
802(1982).
\bibitem{BCFJ}H.Barnum, C.Caves, C.Fuchs, and B.Schumacher,
Phys.Rev.Lett.{\bf 76}, 2818(1996).
\bibitem{BH}V.Bu\v{z}ek, and M.Hillery, Phys.Rev.{\bf A54}, 1844(1996).
\bibitem{BDEF}D.Bru\ss , D.DiVincenzo, A.Ekert, C.A.Fuchs,
C.Macchiavello, and J.A.Smolin, Phys.Rev.{\bf A57}, 2368(1998).
\bibitem{F}C.A.Fuchs, Fortschr.Phys.{\bf 46},535(1998).
\bibitem{BEM}D.Bru\ss , A.Ekert, and C.Macchiavello,
Phys.Rev.Lett.{\bf 81}(1998)2598.
\bibitem{GM}N.Gisin, and S.Massar, Phys.Rev.Lett.{\bf 79},2153(1997).
\bibitem{W}R.F.Werner, Phys.Rev.{\bf A58}, 1827(1998).
\bibitem{G}N.Gisin, Phys.Lett.{\bf A242},1(1998).
\bibitem{BDMS}D.Bru\ss ,G.M.D'Ariano, C.Macchiavello, and
M.F.Sacchi, Phys.Rev.{\bf A62}, 62302(2000).
\bibitem{BCDM}D.Bru\ss , M.Cinchetti, G.M.D'Ariano, and
C.Macchiavello, Phys.Rev.{\bf A62}, 012302(2000).
\bibitem{BHW}V.Bu\v{z}ek, M.Hillery, and R.F.Werner,
Phys.Rev.{\bf A60}, R2626(1999).
\bibitem{BBHB}V.Bu\v{z}ek, S.L.Braunstein, M.Hillery, and
D.Bru\ss , Phys.Rev.{\bf A56}, 3446(1997).
\bibitem{B}V.Bu\v{z}ek, private communication.
\bibitem{KOWY}L.C.Kwek, C.H.Oh, X.B.Wang, and Y.Yeo,
Phys.Rev.{\bf A62}, 052313(2000).
\end{thebibliography}
\end{document}